\documentclass[12pt]{aastex631}   

\usepackage{graphicx}

\def\propdiag{
\begin{figure}
    
     \includegraphics[width=1.0\textwidth]{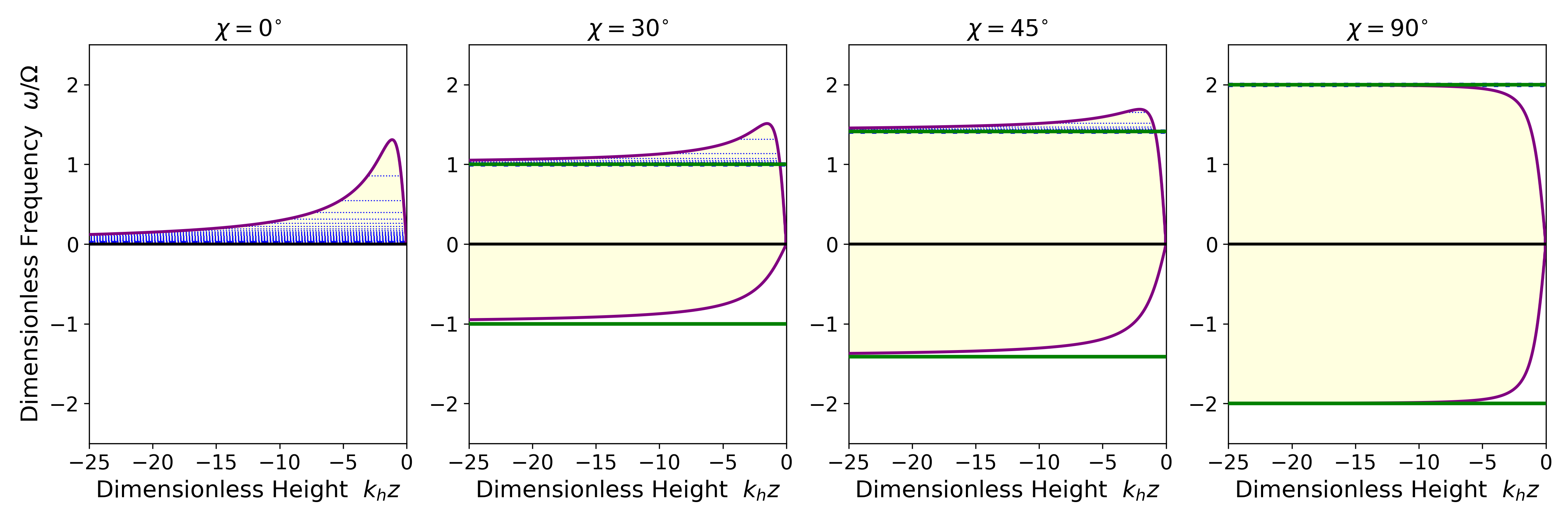}
     \caption{Propagation diagram for a neutrally-stable polytropic atmosphere in a semi-infinite domain. The positive (negative) frequencies indicate waves propagating in prograde (retrograde) direction in the rotating frame. The lightly-shaded region indicates the frequencies of waves that are radially propagating ($k_z^2 > 0$) at the indicated height in the atmosphere. The purple curves denote those frequencies and radii for which a radial turning point occurs (i.e., $k_z^2 = 0$). Thus, at a given frequency, an inertial wave cavity exists in the shaded region between two turning points. Such a cavity for thermal Rossby waves is formed purely by the stratification. We have indicated the eigenfrequencies for such modes for a semi-infinite atmosphere using the blue horizontal dotted lines. The green horizontal lines denote the critical frequencies---see Equation~\eqnref{varpi}---which separate the low-frequency waves that lack a lower turning point from the higher frequency waves that have two turning points (an upper and a lower turning point).}
     \label{fig:figure2}
\end{figure}
}

\def\Eigenfunction{
\begin{figure}
\includegraphics[width=0.5\textwidth]{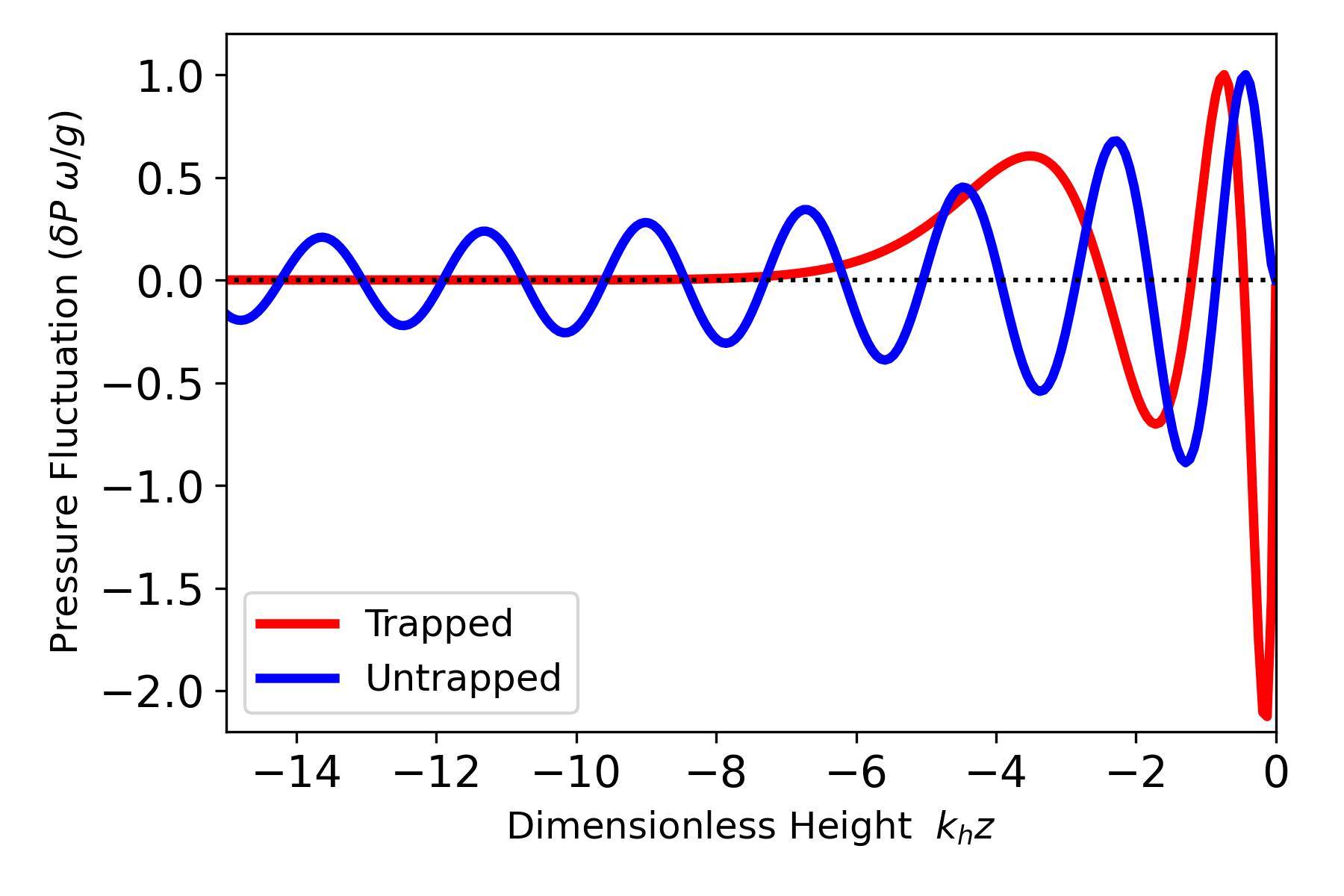}
\caption{Illustration of two types of eigenfunction. The two curves show the normalized Lagrangian pressure fluctuation as a function of dimensionless height for a naturally trapped (red curve) and an untrapped thermal Rossby wave (blue curve). The trapped mode corresponds to a radial order of $n=3$ and has two turning points in radius: an upper turning point that lies very near the upper surface and a lower turning point near $k_h z \approx = -5$. The wave cavity exists between these two turning points and the eigenfunction decays exponentially with depth below the lower turning point. The untrapped wave has only a single turning point. It corresponds to a continuum mode that is launched from infinite depth, travels upwards, reflects off the upper turning point near the origin, and then travels back downwards. This specific continuum mode is prograde with a frequency $\omega/\Omega = 0.5$.  Both eigenfunctions are computed for an angle of propagation of $\chi = 45^{\circ}$.}
\label{fig:eigfuncs}
\end{figure}
}

\def\finitedomain{
\begin{figure}

\includegraphics[width=\textwidth]{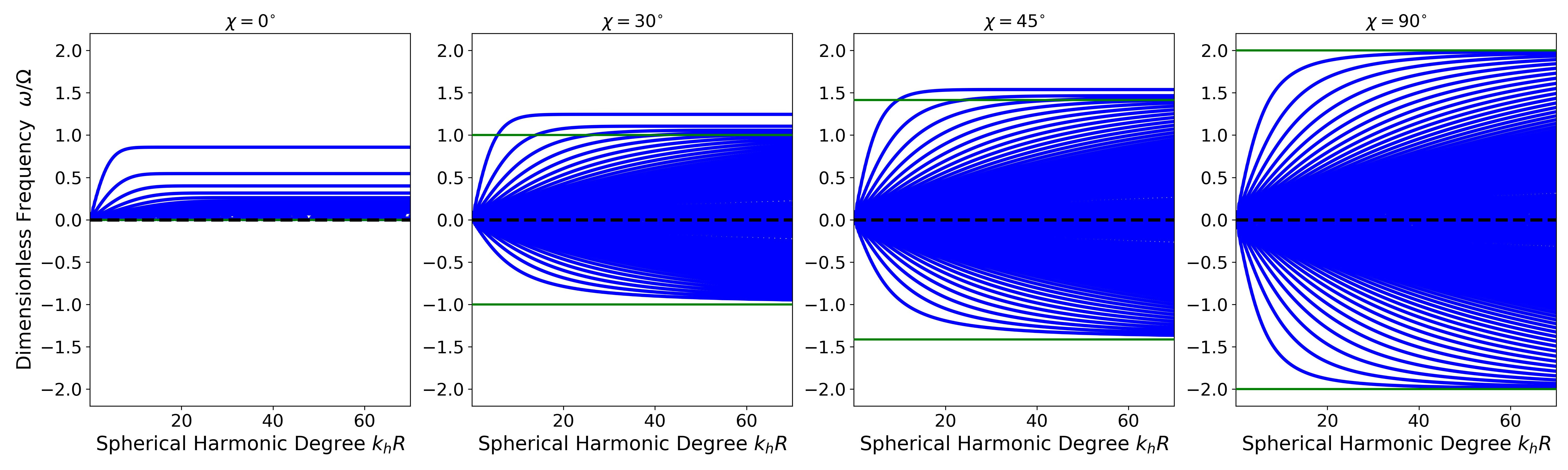}
     \caption{Dimensionless eigenfrequencies as a function of harmonic degree
$k_{h}R$ for a neutrally-stable polytropic atmosphere with a finite depth of $D = 200$ Mm. The green solid line is the same as in Figure~\ref{fig:figure2}. For waves with latitudinal propagation ($\chi > 0$), the uppermost and lowermost curves are for $n=0$ modes lacking radial nodes in their pressure eigenfunctions. Each subsequent radial overtone ($n=1$, $n=2$, and so forth) lies slightly closer to zero frequency. Thus, there is a retrograde and prograde branch for each radial overtone. The solid cone of blue color is artificial and arises because the curves become so close together that they cannot be resolved in the image. The left-most panel, which corresponds to pure zonal propagation, lacks the retrograde lower curves altogether.}
\label{fig:figure7}
\end{figure}
}

\def\millstone{
\begin{figure}
    \includegraphics[width=0.5\textwidth]{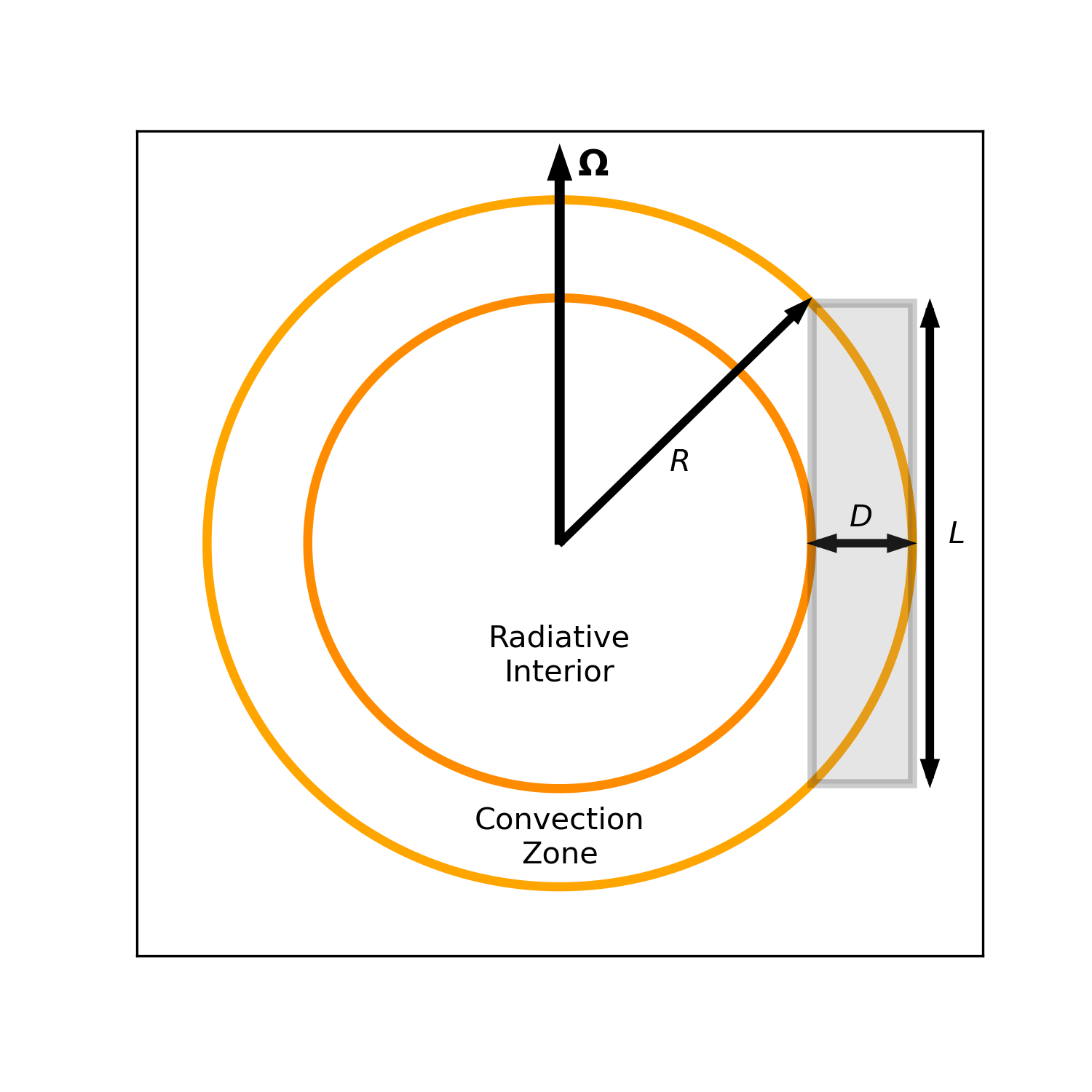}
    \caption{Sketch showing the finite domain considered in Section~\ref{subsec:lat_trapping}. The outer surface of a star and the bottom of the convection zone are indicated in the meridional plane as orange circles. The waves are confined within the convection zone of depth $D$, and within a latitudinal band of length $L$ which is the length of the chord that is tangent to the bottom of the star’s convection zone. The domain is periodic in the azimuthal direction with a circumference length of $2\pi R$. A meridional slice through the domain is shown as the gray box.}
    \label{fig:millstone}
\end{figure}
}

\def\latovertone{
\begin{figure}
\includegraphics[width=0.5\textwidth]{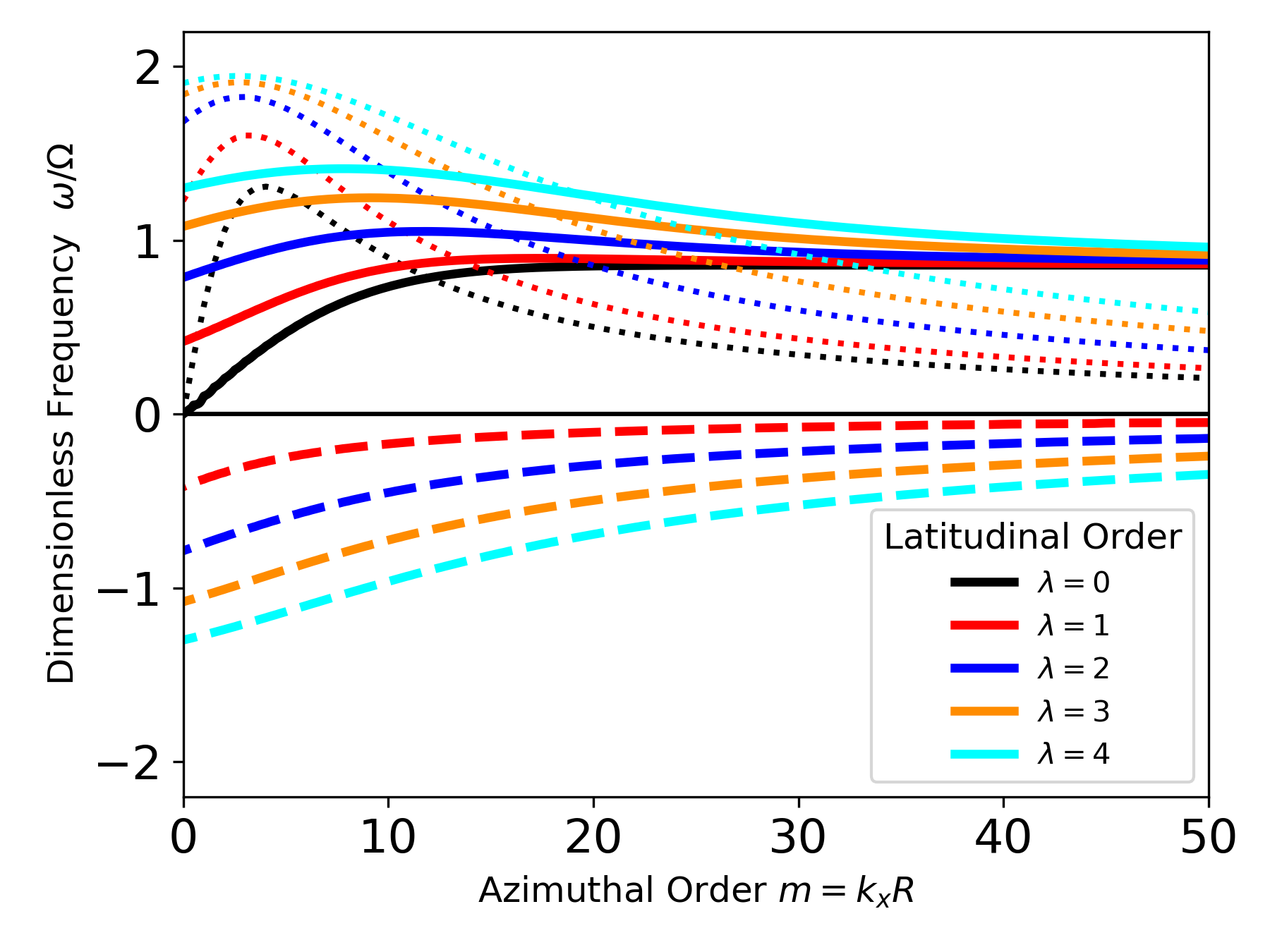}
\caption{Eigenfrequencies of modes trapped latitudinally and radially in the millstone domain illustrate in Figure~\ref{fig:millstone}. The frequencies are plotted as a function of the azimuthal order. Each distinct color corresponds to a different latitudinal order $\lambda$, as indicated in the legend. The solid curves show the prograde-propagating thermal Rossby waves, whereas the dashed curves are for the retrograde inertial waves. The dotted curves indicate the frequency where the lower turning point passes through the bottom boundary. Wave frequencies above the appropriate dotted curve have two turning points in the radial domain and have a wave cavity that is screened from the lower boundary condition.  Frequencies below the dotted curve possess only an upper turning point in the domain and are reflected off the lower boundary.}
\label{fig:lat_overtones}
\end{figure}
}

\usepackage{natbib}

\newcommand{\bvec}[1]{{\mbox{{\boldmath$#1$}}}}	
\newcommand{\unitv}[1]{\bvec{\hat{#1}}}			
\newcommand{\grad}{\bvec{\nabla}}			    

\newcommand{\eqnref}[1]{(\ref{#1})}

\begin{document}

\title{Latitudinal Propagation of Thermal Rossby Waves in Stellar Convection Zones}

\author{Rekha Jain}
\affil{School of Mathematics and Statistics, University of Sheffield, S3 7RH UK}
\email{R.Jain@sheffield.ac.uk}

\author{Bradley W. Hindman}
\affil{JILA, University of Colorado, Boulder, CO~80309-0440, USA}
\affil{Department of Applied Mathematics, University of Colorado, Boulder, CO~80309-0526, USA}

\begin{abstract}
Using an analytic model, we derive the eigenfrequencies for thermal Rossby waves
that are trapped radially and latitudinally in an isentropically stratified atmosphere.
We ignore the star's curvature and work in an equatorial f-plane geometry.
The propagation of inertial waves is found to be sensitive to the relative direction
of the wave vector to the zonal direction. Prograde propagating thermal Rossby waves are 
naturally trapped in the radial direction for frequencies above a critical threshold, which 
depends on the angle of propagation. Below the threshold frequency, there exists a 
continuous spectrum of prograde and retrograde inertial waves that are untrapped in an isentropic atmosphere, but can be trapped by gradients in the specific entropy density such as occurs in a stellar convection zone. Finally, we discuss the implications of these waves on recent observations of inertial oscillations in the Sun, as well as in numerical simulations. 
\end{abstract}


\section{Introduction}
A primary motivation for studying inertial oscillations of stars is their implications in understanding the stellar interior structure. In particular, observations of such oscillations may provide a strong constraint on  superadiabaticity and other thermodynamic variables within the star's convection zone \citep[e.g.,][]{Gilman1987}. Recent observations in the Sun of Rossby waves and other inertial oscillations \citep[e.g.,][]{Loeptien2018, Hanasoge2019, Proxauf2020, Gizon2021, Hathaway2021, Hanson2022} have aroused interest in using these waves as seismic probes of the solar interior and in the potential role that they play in the Sun's magnetic cycle \citep[e.g.,][]{Dikpati2020}. Through exploitation of the Sun's acoustic oscillations ($p$ modes), helioseismology has successfully mapped the Sun’s differential rotation and its thermal structure throughout the convection zone \citep[see][]{Christensen2002}. However, some quantities, such as the turbulent viscosity and radial entropy gradient in the convection zone, are essentially invisible to the $p$ mode oscillations. Further, the $p$ modes, as measured from the ecliptic, do not sample high-latitudes well; therefore, flows and thermal structures in the polar caps are poorly constrained. All of these missing pieces are important elements in theories of stellar dynamics and the dynamo. Thus, there remains a prominent gap in our understanding of the solar interior. The observation of these new class of oscillations, which are likely to have sensitivity to many of these parameters, could bridge that gap.

Thermal Rossby waves \citep[e.g.,][]{Roberts1968, Busse1970, Hindman2022} are a specific type of gravito-inertial wave that corresponds to convective modes that have been partially or completely stabilized by rotation.  While they have yet to be detected in the Sun observationally, they are ubiquitous in laboratory experiments of convection in a rotating fluid \citep[e.g.,][]{Mason1975, Busse1982, Smith2014, Lonner2022}. Further, thermal Rossby waves are seen in numerical simulations of stellar convection and are a crucial ingredient in the maintenance of a star's differential rotation \citep[e.g.,][]{Brun2011, Miesch2000, Hindman2020}. These waves appear at convective onset and persist even when the fluid becomes turbulent. In a Boussinesq fluid, the waves fill the spatial domain; but, in a gravitationally stratified fluid, numerical simulations have shown that the waves can be trapped in radius, being concentrated near the surface or near the bottom of the convection zone depending on the stratification \citep{Jones2009, Hindman2020, Hindman2023}. This strong variation in the location of the wave cavity is a clear indication that these waves are extremely sensitive to the stratification, and in particular the radial entropy gradient. Thus, if detected, such waves would serve as an excellent seismic probe.

While there is a rich literature on Rossby waves in stratified astrophysical disks \citep[e.g.,][]{Li2000, Lin2012}, to date only a few studies have explored how thermal Rossby waves propagate through a stratified star \citep{Glatzmaier1981, Gilman1987, Hindman2022, Hindman2023}. Further, for simplicity, all of these have ignored propagation and reflection in the latitudinal direction. Studying the radial and latitudinal wave cavity of low-mass stars with a near-surface convection zone will be our main aim here. The present paper considers the propagation of thermal Rossby waves and their kin in all three directions (zonal, latitudinal and radial) within an adiabatically stratified background atmosphere. 
Section 2 describes a model to this effect and in Section 3 we derive the resulting governing
equation and discuss the nature of the solutions. In section 4, we consider the eigenmodes of a semi-infinite polytropic atmosphere and in Section 5 we consider finite domains. Finally, we summarize and present our conclusions in Section 6.

\section{The Model}

Although most stars are nearly spherical, we consider a local Cartesian coordinate system by defining a tangent plane at the star's equator
and assuming that the rotation vector is uniform over the whole tangent
plane. This f-plane approximation simplifies the study of waves that
have short horizontal wavelengths. Therefore, we place the origin at the
stellar surface with the unit vectors $\unitv{x}$, $\unitv{y}$ and
$\unitv{z}$ pointing in the longitudinal, latitudinal and radial
directions, respectively. We adopt uniform rotation to avoid singularities in the equations resulting from critical layers where the local rotation rate equals a wave's phase speed \citep[e.g.,][]{Gilman1987, Gizon2021}.

We investigate the linearized fluid equations in the
absence of a background flow. We consider a steady-state background,
denoted with subscript $0$, with perturbations about that background indicated with subscript
$1$.  The background atmosphere is assumed to be a plane-parallel
atmosphere which is gravitationally stratified with a constant
gravitational acceleration $\bvec{g} = - g{\unitv{z}}$. Thus, the
background pressure and density vary as a function of $z$
and are denoted by $P_{0}(z)$ and $\rho_{0}(z)$,
respectively. These quantities are related by hydrostatic
balance and the ideal gas law.
The linearized equation of motion that governs a rotating, inviscid fluid on an f-plane is given by:

\begin{eqnarray}
\label{momentum}
\frac{\partial{{\bvec{u}}_{1}}}{\partial t} = - 2 \left(\bvec{\Omega} \times {\bvec{u}_{1}}\right) -\frac{1}{\rho_{0}}\grad P_{1} - \frac{g\rho_{1}}{\rho_{0}}{\unitv{z}} \;,
\end{eqnarray}

\noindent where ${\bvec{u}}_{1} = u\unitv{x} + v\unitv{y} +
w\unitv{z}$ is the perturbed fluid velocity and $P_1$ and $\rho_1$ are the Eulerian fluctuations of the pressure and density. The rotation vector
points purely in the latitudinal direction, $\bvec{\Omega} = \Omega
\unitv{y}$. The linearized continuity equation for a compressible fluid is,

\begin{eqnarray}
\label{continuity}
\frac{\partial \rho_{1}}{\partial t} + \rho_{0} (\grad \cdot \bvec{u}_1) - \rho_{0}\frac {w}{H} = 0 \; ,
\end{eqnarray}

\noindent where

\begin{eqnarray*}
H= - \left(\frac{d \ln\rho_{0}}{dz}\right)^{-1} \; ,
\end{eqnarray*}

\noindent is the density scale height. We consider adiabatic motions such that

\begin{eqnarray}
\label{adiabatic}
\frac{\partial \delta P}{\partial t} = \frac{\partial P_{1}}{\partial t} - g\rho_{0} w = c_{s}^{2}
\left(\frac{\partial \rho_{1}}{\partial t} + w \frac{d \rho_{0}}{dz} \right)
\end{eqnarray}

\noindent where $\delta P$ is the Lagrangian pressure fluctuation and $c_{s}^{2}$ is the square of the sound speed defined by

\begin{eqnarray*}
c_{s}^{2}(z) = \frac{\gamma P_{0}}{\rho_{0}}
\end{eqnarray*}

\noindent with $\gamma$ as the ratio of specific heats.

\section{The Governing Wave Equation}
\label{sec:maths} 

Since the background atmosphere is steady and invariant in the $\unitv{x}$ and
$\unitv{y}$ directions, we seek horizontal plane-wave solutions of the form

\begin{equation}
    f(x,y,z,t) = \tilde{f}(z) \, e^{i(k_{x}x+k_{y}y)} \, e^{-i\omega t} \; ,
\end{equation}

\noindent where $f$ is any perturbed fluid variable. Furthermore, $k_{x}$ and $k_{y}$ are the wavenumbers in the $x$ and
$y$ directions and $\omega$ is the temporal frequency. Without loss of generality, we only consider positive longitudinal wavenumbers, $k_{x} > 0$. The waves
propagate in the prograde direction if the frequency is positive,
$\omega > 0$, and in the retrograde direction for negative frequencies,
$\omega < 0$.

Using the above plane-wave form along with equations
(\ref{momentum})--(\ref{adiabatic}), we obtain the following governing equation for $\delta P$
\begin{eqnarray}
    \label{governing}
    \left\{\frac{d^{2}}{dz^{2}} + \frac{1}{H}\frac{d}{dz} + \Lambda^2(z) \right\}\delta P = 0 \; , \qquad\qquad
\\
    \nonumber
    \Lambda^2(z) \equiv\frac{\omega^{2} - 4  \Omega^{2}}{c_{s}^{2}} - k_{h}^{2}  \left(1 - 
\frac{N^{2}}{ \omega^{2}} \right) 
  +\frac{2  \Omega k_{x}}{ \omega \cal H}  +  \frac{4
k_{y}^{2}\Omega^{2}}{\omega^{2}} \;.
\end{eqnarray}



\noindent In this Equation, $k_h$ is the total horizontal wavenumber, i.e., $k_h^2 = k_x^2 + k_y^2$. Further, the square of the buoyancy frequency, $N^2$, is defined as follows
\begin{eqnarray}
N^{2}(z) = {g}\left( \frac{1}{H} - \frac{g}{c_{s}^{2}} \right),
\end{eqnarray}

\noindent and ${\cal H}$ is a scale height that depends on $H$ and $N^2$
  
\begin{eqnarray}
\label{scale}
\frac{1}{\cal H} = \left( \frac{1}{H} - \frac{2N^{2}}{g} \right).
\end{eqnarray}

The governing wave equation (\ref{governing}) can be written as a Helmholtz equation through the substitution $\delta P =\sqrt{\rho_{0}} \, \Psi(z)$
,
\begin{eqnarray}
\label{Helmholtz}
\frac{d^{2} \Psi}{dz^{2}}  + k_{z}^{2}\,\Psi = 0,
\end{eqnarray}

\noindent with

\begin{eqnarray}
\label{local}
k_{z}^{2} = \frac{\omega^{2} - (\omega_{ac}^{2}+4\Omega^{2})
}{c_{s}^{2}} - k_{h}^{2}  \left(1 -  \frac{N^{2}}{ \omega^{2}} \right) 
+  \frac{2  \Omega k_{x}}{ \omega \cal H} + \frac{4
k_{y}^{2}\Omega^{2}}{\omega^{2}} .
\end{eqnarray}
\
\\
Here,
\begin{eqnarray}
\label{gen_acoustic_cut_off}
\omega_{ac}^2 \equiv \frac{c_s^2}{4H^2} \left(1-2\frac{dH}{dz}\right)
\end{eqnarray}
\noindent is the square of the acoustic cut-off frequency, $\omega_{ac}$.

Equation~\eqnref{local} provides a local dispersion relation for both acoustic waves and gravito-inertial waves. In the {\bf {low-frequency limit}}, Equation (\ref{local}) reduces to a local dispersion relation for just gravito-inertial waves,

\begin{eqnarray}
\label{low-frequency}
k_{z}^{2} = \left [\frac{2\Omega k_{x}}{\omega {\cal H}} + \frac{4
k_{y}^{2}\Omega^{2}}{\omega^{2}} +
k_{h}^{2}\frac{N^{2}}{\omega^{2}} \right]-  \left(k_{h}^{2} +
\frac{\omega^{2}_{ac}}{c_{s}^{2}} \right) \; .
\end{eqnarray}

\noindent The two terms in the parentheses provide a negative contribution and lead to vertical evanescence (i.e., $k_z^2 < 0$). Conversely, the terms inside the square brackets can be positive, thereby leading to vertical propagation ($k_z^2 >0$). The first two terms in the brackets arise from the Coriolis force. As discussed in \citet{Hindman2022}, the first of these terms is positive for prograde waves and can produce propagating thermal Rossby waves.  The second is a newly identified term that also leads to vertical propagation and is responsible for the axisymmetric inertial waves in a sphere previously studied by \citet{Guenther1985}. As we will see in subsequent sections, this term can in fact lead to vertical detrapping for waves of very low frequency. Finally, the third term in the square brackets is the buoyancy term responsible for internal gravity waves.

\subsection{Neutrally Stable Atmosphere}

Previous studies of latitudinal propagation in a compressible atmosphere have been carried out by \citet{Thuburn2002} and \citet{Kasahara2003}; but these efforts considered a stably-stratified isothermal atmosphere and investigated the modifications to acoustic-gravity waves by the Coriolis force. Further, since the atmosphere was isothermal, waves cannot be naturally trapped by the stratification. Thus, it is important to explore the radial and latitudinal propagation in an atmosphere where the density scale height varies with height. \citet{Hindman2022} demonstrated that inertial waves in a neutrally stable convection zone can be trapped in the radial direction. In that paper we considered waves that did {\bf not} propagate latitudinally.  Here we demonstrate that radial trapping is still possible when latitudinal propagation is allowed, but there is also a continuous spectrum of extremely low-frequency inertial waves that are free to propagate to any depth.

We now consider an isentropic atmosphere such that $N^{2} = 0$, i.e., the buoyancy forces disappear and the Coriolis force is the only restoring force for the low-frequency waves.
Such a neutrally stable atmosphere is polytropic and possesses a single height at which the pressure, density, and temperature all vanish. We place the origin $z=0$ at this singular point and let the atmosphere exist within the region below, for $z < 0$.  In an isentropic polytrope the atmospheric profiles have the following power-law forms:
\begin{eqnarray}
    \label{rhoprofile}
    \rho_0(z) &=& A_0 (-z)^\alpha \; , 
\\
    P_0(z) &=& \frac{g A_0}{\alpha + 1} (-z)^{\alpha+1} \; ,
\\
    c_s^2(z) &=& \frac{\gamma g}{\alpha+1}(-z) \; ,
\\
    \label{Hprofile}
    H(z) = {\cal H}(z) &=& \frac{(-z)}{\alpha} \; ,
\\
    \label{ACprofile}
    \frac{\omega_{ac}^2(z)}{c_s^2(z)} &=& \frac{\alpha(\alpha+2)}{z^2}
\end{eqnarray}

\noindent where $A_0$ is an arbitrary scale factor and the dimensionless parameter $\alpha$ is the polytropic
index given by $\alpha = \left(\gamma - 1\right)^{-1}$.

\subsection{Propagation Diagram and Eigenmodes}
For an isentropic stratification, where $N^2=0$, the low-frequency form of the local dispersion relation~\eqnref{low-frequency} reduces to

\begin{eqnarray}
\label{lowfreq}
k_{z}^{2}= \left[\frac{2 \alpha \Omega k_{x}}{ \omega (-z)} + 4 k_{y}^{2} \frac{\Omega^2}{\omega^2}\right] - \left(k_h^{2} +
\frac{\alpha(\alpha + 2)}{4z^{2}}\right) \; .
\end{eqnarray}

\noindent Near the upper boundary of the atmosphere ($z\to 0$), the second term in the parentheses (arising from the acoustic cut-off frequency) is large and leads to reflection and vertical evanescence. Deep within the atmosphere ($z \to -\infty$), the dispersion relation reduces to
\begin{eqnarray}
    \label{deep_propagation}
    k_z^2 \approx k_h^2 \left(\varpi^2-1\right) \; ,
\end{eqnarray}

\noindent where, for later convenience, we have defined

\begin{eqnarray}
\label{varpi}
    \varpi^2 \equiv \frac{k_y^2}{k_h^2} \frac{4\Omega^2}{\omega^2} = \frac{4\Omega^2}{\omega^2} \sin^2\chi\; .
\end{eqnarray}

\noindent Here, $\chi$ indicates the direction of horizontal propagation, with $\chi = 0$ corresponding to pure prograde propagation and $\chi = \pi/2$ indicating pure northward propagation ($k_x = k_h \cos\chi$ and $k_y = k_h \sin\chi$). From Equation~\eqnref{deep_propagation}, we can easily determine that the inertial waves can be either vertically evanescent ($k_{z}^{2} < 0$) or vertically propagating ($k_{z}^{2} > 0$) depending on the frequency and the horizontal direction of propagation. A wave cavity exists when the waves become evanescent, which requires $\varpi^2 < 1$ or equivalently,
\begin{eqnarray}
    \label{omega_crit}
    \omega > 2\frac{|k_y|}{k_h}\Omega = 2\Omega |\sin\chi| \; .
\end{eqnarray}

\noindent These waves are naturally trapped by the density stratification and form a discrete spectrum of inertial eigenmodes that in the limit of $k_y=0$ become the thermal Rossby waves of \citet{Hindman2022} and the fast branch of thermal Rossby waves as discussed by \citet{Hindman2023}. For frequencies lower than the critical value given in Equation~\eqnref{omega_crit}, the waves remain vertically propagating to all depths and a downward propagating inertial wave is never reflected back upwards in a semi-infinite domain: no cavity exists. This family of solution corresponds to a continuous spectrum of untrapped inertial eigenmodes.

This behavior is fully revealed in Figure~\ref{fig:figure2} which provides a propagation diagram as a function of height within an isentropic atmosphere. The lightly shaded region indicates those frequencies that correspond to vertically propagating waves. The green horizontal lines indicate the critical value (and its negative) provided by Equation~(\ref{omega_crit}). The frequencies above the upper bound correspond to trapped inertial waves, while frequencies between the two bounds constitute the untrapped modes. As one can see, the trapped waves propagate between two turning points and therefore form a discrete spectrum of normal modes. Conversely, the untrapped inertial waves are screened from the origin by an upper turning point, but lack a lower turning point. The trapped waves all possess positive frequencies, and are hence prograde-propagating, whereas the untrapped continuum modes can be prograde thermal Rossby waves or retrograde inertial waves.

\subsection{Vertical Wave Equation for an Isentropic Stratification}

For the polytropic atmosphere, the Helmholtz equation with $k_{z}^{2}$ given by Equation
(\ref{lowfreq}) can be transformed into the well-known Whittaker
Equation by making a change of variable $\zeta =- 2 \sqrt{1-\varpi^2} \, k_h z$,

\begin{eqnarray}
\label{Whittaker}
\frac{d^{2}\Psi}{d \zeta^{2}} + \left[ \frac{\kappa}{\zeta} - \frac{1}{4} +
\frac{1/4-\mu^{2}}{\zeta^{2}}\right] \Psi = 0
\end{eqnarray}

\noindent where $\mu \equiv (\alpha + 1)/2$ is a constant. The
parameter $\kappa$ is the eigenvalue of the second-order ordinary
differential equation (\ref{Whittaker}) and it depends on the frequency
and direction of propagation of the wave, i.e.

\begin{eqnarray}
\label{eigenvalue}
\kappa = \frac{\alpha \Omega}{\omega}\frac{\cos\chi}{  \sqrt{1-\varpi^2}} \; .
\end{eqnarray}

\noindent Notice, that for all of the trapped prograde waves (refer to Figure~2) we have $\varpi^2 < 1$; hence, for these trapped waves, the parameter $\kappa$ and the dimensionless depth $\zeta$ have real values. Conversely, the low-frequency waves without a lower turning point have $\varpi^2 >1$ and both $\kappa$ and the dimensionless depth $\zeta$ are purely imaginary.

Whittaker's Equation (\ref{Whittaker}) has two solutions, the Whittaker functions $M_{ \kappa  \mu}( \zeta)$ and $W_{ \kappa  \mu}(\zeta)$ \citep[see][]{Abramowitz1968}. These can be expressed with Kummer's
confluent hypergeometric functions $M$ and $U$ as follows:
\begin{eqnarray*}
M_{ \kappa  \mu}( \zeta) = e^{- \zeta/2}  \zeta^{ \mu+ \frac{1}{2}} M \left(-
\eta, 1+2 \mu,  \zeta \right)
\\
W_{ \kappa  \mu}( \zeta) = e^{- \zeta/2}  \zeta^{ \mu+\frac{1}{2}} U (- \eta,
1+2 \mu,  \zeta)
\end{eqnarray*}

\noindent with
\begin{eqnarray}
  \eta \equiv \kappa - \left(\mu + \frac{1}{2}\right) \; .  
\end{eqnarray}

\section{Semi-Infinite Domain}

In this section we consider an atmosphere of semi-infinite extent in radius.  The atmosphere has a physical upper surface (at $z=0$), but is infinitely deep ($z\to -\infty$). Such an atmosphere is valid to use if the waves are naturally trapped and are confined over a region of finite radius. However, here we primarily explore the eigenmodes for such an atmosphere because their mathematics provides illumination for the behavior of waves in finite domains.

\subsection{Naturally Trapped Modes in Radius}

The radially trapped modes are acquired by demanding that the solutions vanish and remain regular at the two singular points of Whittaker's Equation (\ref{Whittaker}), $\zeta = 0$ and $\zeta  \rightarrow  \infty$,
 \begin{eqnarray}
\delta P(z) = \rho_{0}^{1/2} \Psi(z) =  C_n \, \zeta \, e^{-\zeta/2} \, M (-n,  \alpha+2, \zeta) \; ,
\end{eqnarray}

\noindent where $C_{n}$ is an arbitrary constant and the parameter $\eta$ must be a non-negative integer, $\eta = n \in[0, 1, 2, 3, ...]$, in order to avoid divergence of the eigenfunction in the limit $z\to -\infty$. Therefore, the eigenvalue $\kappa$ takes on discrete values that depend on the radial order $n$ and the polytropic index $\alpha$,
\begin{eqnarray} 
\kappa_{n} = n + 1+ \frac{\alpha}{2} \; .
\end{eqnarray}

\noindent Since the frequency depends on the eigenvalue through Equation~(\ref{eigenvalue}), the discretization of the eigenvalue leads to discretization of the corresponding frequencies.  These eigenfrequencies depend on the direction of propagation $\chi$, the radial order $n$, the rotation rate $\Omega$, and the polytropic index $\alpha$,
\begin{eqnarray}
\label{eigenfre}
  \frac{\omega_{n}}{\Omega} = \left[
    \frac{\alpha^2\cos^2\chi}{(n+1+\alpha/2)^2} + 4\sin^2\chi 
    \right]^{1/2} \;.
\end{eqnarray}

\noindent The blue dotted horizontal lines in Figure~\ref{fig:figure2} indicate these discrete eigenfrequencies. There are a countable infinity of such modes with an accumulation point at the critical frequency, i.e., $\omega \to 2\Omega \left|\sin\chi\right|$ as $n\to \infty$.

The expression for the eigenfrequency, $\omega_{n}$, in Equation~(\ref{eigenfre}) clearly shows that the eigenfrequencies are unaffected by the magnitude of the horizontal wavenumber, $k_h$, even though they do depend on the direction of propagation $\chi$. This is a consequence of the self-similarity of a polytropic atmosphere \citep[for details, see][]{Hindman2022}. For purely northward or southward propagation, i.e., $\chi = \pm \pi/2$, the eigenfrequencies become independent of the radial order $n$ and on the stratification $\alpha$ taking on the value $\omega_{n} = 2\Omega$. We only illustrate positive values of $\chi$ because the eigenfrequencies are symmetric in $\chi$, i.e., $\omega_n(-\chi) = \omega_n(\chi)$.  

The lower turning point can be calculated directly from Equation~(\ref{lowfreq}) by solving for the two heights in the atmosphere where $k_z^2 = 0$. The deepest of these two solutions corresponds to the lower turning point,
\begin{eqnarray}
    -k_h z_{\rm turn} = \frac{\kappa_n + \sqrt{\kappa_n^2 -\alpha(\alpha+2)/4}}{\sqrt{1-\varpi^2}}.
\end{eqnarray}

\noindent Note that the lower turning point for all radial orders deepens as the propagation angle increases from zero.  Hence, the cavity is shallowest for purely zonal propagation ($\varpi^2 = 0$) and becomes infinitely deep for purely latitudinal propagation ($\varpi^2 = 1$).  Below the lower turning point, the eigenfunction becomes evanescent and decays exponentially with depth.  This behavior is illustrated in Figure~\ref{fig:eigfuncs}.  The red curve illustrates the eigenfunction for a $n=3$ mode with a lower turning point near $k_h z= -5$.

\subsection{Untrapped Modes in Radius}

Waves with frequencies that lie with the band
\begin{eqnarray}
    \label{continuous_spectrum}
    |\omega| < 2 \Omega  |\sin\chi| \; ,
\end{eqnarray}
do not possess a lower turning point.  Hence, these waves continue to propagate deep in the atmosphere. This behavior leads to a continuous spectrum of wave modes that are regular at the origin, $z=0$, and oscillatory in the limit $z\to -\infty$. The regularity condition at the origin is the only boundary condition that we can physically enforce. This leads to eigenfunctions of the form,
\begin{eqnarray}
    \label{untrapped}
    \delta P = C(\omega) \, \zeta \, e^{-\zeta/2} M\left(\mu + 1/2 - \kappa, \alpha + 2, \zeta\right) \; ,
\end{eqnarray}

\noindent where $C(\omega)$ is an amplitude determined by initial conditions. The dimensionless depth $\zeta$ and the eigenvalue $\kappa$ are both imaginary and $\kappa$ is no longer discrete,
\begin{eqnarray}
    \kappa = -i \frac{\alpha\Omega}{\omega} \frac{\cos\chi}{\sqrt{\varpi^2 -1}} \; ,
\\
    \zeta = -2i \sqrt{\varpi^2-1} \, k_h z \; . 
\end{eqnarray}

\noindent The eigenfunction given by Equation~(\ref{untrapped}) is a standing wave that is comprised of a wave launched from infinite depth that travels upwards, reflects off the acoustic cut-off frequency near the origin, and then travels downwards back to infinite depth.  Such an eigenfunction is illustrated in Figure~\ref{fig:eigfuncs}, where the blue curve shows a continuum mode that lacks a lower turning point.  The wave remains oscillatory in the limit $z\to-\infty$.

For special integer values of the polytropic index, $\alpha = 2L$, where $L$ is any nonnegative integer, solutions for the untrapped continuum modes can be written in terms of real functions with real parameters and arguments using the regular Coulomb wave function $F_L$ \citep[see][]{Abramowitz1968},

\begin{equation}
    \delta P = C(\omega) \, z^{\alpha/2} F_L\left(q, {\cal Z}\right) \; ,
\end{equation}

\noindent with

\begin{eqnarray}
    q &\equiv& -i\kappa = -\frac{\alpha \Omega}{\omega} \frac{\cos\chi}{\sqrt{\varpi^2-1}} \; ,
\\
    {\cal Z} &\equiv& -\frac{i}{2} \zeta = -\sqrt{\varpi^2-1} \, k_h z \; .
\end{eqnarray}

\section{Finite Domain}
\subsection{Radial Trapping}
In a solar-like star, the convection zone is approximately 200 Mm deep with the stably-stratified radiative zone lying underneath. Within the transition between the two layers, the buoyancy frequency jumps dramatically and this large change should make the bottom of the convection zone an efficient reflector of gravito-inertial waves.  Therefore, an appropriate model of inertial waves in a star's convection zone is to apply a regularity condition at the origin (as we did in the previous section) and a reflective
lower boundary condition at a finite depth of $D$.  For this later condition we adopt $\delta P(z=-D) = 0$ for $D=200$ Mm. The global dispersion relation satisfying these boundary conditions is
\begin{eqnarray}
    \label{finite}
    M\left(\mu+1/2-\kappa, \, \alpha + 2, \, 2 \sqrt{1-\varpi^2} k_h D \right) = 0 \; . 
\end{eqnarray}

\noindent We note that the imposition of a lower boundary condition at a finite depth converts the continuous spectrum of untrapped waves into a discrete spectrum. In Figure~\ref{fig:figure7}, the dimensionless frequency $\omega/\Omega$ is plotted as a function of $k_{h}R$ for four different angles of propagation. Note that $k_{h}R$ corresponds to the harmonic degree of spherical harmonics in a spherical geometry. In each panel, the uppermost curve with the highest positive frequencies corresponds to modes that lack nodes in radius (i.e., radial order $n=0$). The second highest indicates modes with one radial node ($n=1$). Sequentially lower curves have one additional node and have an accumulation point at zero frequency for an infinite number of nodes.  All of these modes have positive frequencies and are prograde propagating. The curve with the largest negative frequency also lacks radial nodes ($n=0$), but corresponds to retrograde-propagating inertial waves. Each subsequent curve with smaller negative frequencies has an additional node with an accumulation point at zero frequency as well.  The horizontal green lines indicate the frequency bounds that separate the modes that are naturally trapped by the stratification ($\omega > 2\Omega\left|\sin\chi\right|$) from those that would have been untrapped but are now trapped by the lower boundary of the finite domain.

\subsection{Latitudinal Trapping}
\label{subsec:lat_trapping}

Since we have derived the inertial waves within an equatorial f-plane model instead of spherical geometry, we have implicitly assumed that the waves are confined near the equator and have short horizontal wavelengths, i.e., $k_h R \gg 1$ where $R$ is the star's photospheric radius.  The first of these assumptions allows us to ignore the curvature terms in the fluid equations that arise from the spherical geometry. The second can be justified by examining the results of numerical simulations and eigenmode calculations in spherical geometry \citep[i.e.,][]{Jones2009, Hindman2020, Bekki2022b}, which clearly indicate that thermal Rossby waves are indeed confined or trapped near the equator.

The traditional way to capture latitudinal trapping in plane-parallel geometry is to adopt an equatorial $\beta$-plane approximation, where all atmospheric and geometric terms in the fluid equations are linearized with respect to the latitudinal coordinate (i.e., one assumes $y/R \ll 1$). We will not do so here to avoid the resulting complication of solving the wave equations in a truly 2D atmosphere.  Instead, we will retain our f-plane geometry but make the simple assumption that the waves are confined within a latitudinal band that extends north and south of the equator by a fixed distance $L/2$. To enforce reflection at $y = \pm L/2$, we impose Neumann boundary conditions on the Lagrangian pressure fluctuation,

\begin{equation}
    \left.\frac{\partial \delta P}{\partial y}\right|_{y = \pm L/2} = 0 \; ,
\end{equation}

\noindent which is equivalent to an impenetrable boundary condition ($v=0$). Such boundary conditions are quite similar to those employed in the study of Rossby waves in astrophysical disks \citep{Lin2012}.

To simplify comparison with waves in a spherical geometry, we will assume that the longitudinal direction is periodic which quantizes the longitudinal wavenumber, $k_x = m / R$, with $m$ being the azimuthal order of the concomitant spherical harmonic. Thus, our spatial domain is shaped like a millstone, with an outer annular radius of $R$, an inner radius of $R-D$, and a cylindrical height of $L$ (see Figure~\ref{fig:millstone}). In numerical simulations, the latitudinal extent of thermal Rossby wave eigenfunction varies as a function of horizontal wavenumber \citep[i.e.,][]{Hindman2020}. But, in general, the waves often fill the region outside the cylinder that is tangent to the base of the convection zone at the equator. Hence, we choose the width of the latitudinal band $L$ to be the length of the chord that is tangent to the bottom of the star's convection zone (see Figure~\ref{fig:millstone}),

\begin{equation}
    L = 2\sqrt{R^2 - (R-D)^2} \; .
\end{equation}

\noindent Using $R= 700$ Mm and $D = 200$ Mm, one obtains $L \approx 980$ Mm.  

The latitudinal boundary conditions discretizes the latitudinal wavenumber,

\begin{equation}
    k_y = \frac{\lambda \pi}{L}\; , \quad  \lambda = \left[0,1,2,3, ...\right] \; ,
\end{equation}

\noindent leading to eigenfunctions of the form,

\begin{equation}
\label{millstone_eig}
    \delta P(x,y,z,t) = C \cos\left[\lambda (y-L/2)\right] \, e^{im x/R} \, \zeta e^{-\zeta/2} \, M(-\eta, \alpha + 2, \zeta) \, e^{-i\omega t} \; .
\end{equation}

\noindent The quantum number $\lambda$ can be any non-negative integer, with the value of $\lambda$ indicating the number of latitudinal nodes that appear in the eigenfunction. Modes with $\lambda = 0$ correspond to $k_y = 0$ and we have explored these modes previously in \citet{Hindman2022}. We recognize that our choice of $L$ results in large latitudinal wavelengths for low latitudinal orders (small $\lambda)$; hence, they break the short wavelength approximation. Our goal, however, is not to derive accurate quantitative frequencies but to instead generate a general qualitative understanding of the wave behavior. Hence, we carry on nonetheless.

The eigenfrequencies for this ``millstone" model can be generated by solving Equation~\eqnref{millstone_eig} numerically, for the discrete values of $k_x$ and $k_y$ that we just discussed.  Figure~\ref{fig:lat_overtones} presents the results. The eigenfrequencies are plotted as a function of azimuthal order $m$ for the five lowest latitudinal orders. For clarity of presentation, only the radial fundamental modes, lacking radial nodes ($n=0$), are illustrated. The solid curves show the prograde-propagating thermal Rossby waves and the dashed lines correspond to the retrograde-propagating inertial waves. The color of the curve indicates the latitudinal order. The reader should note that for the $\lambda=0$ mode, which propagates purely zonally ($k_y = 0$), the retrograde solution is missing because it becomes a zero-frequency geostrophic mode \citep{Hindman2022}. The dotted curves show the frequency for which the lower turning point passes through the bottom boundary on the convection zone. Every frequency, for the appropriate value of $\lambda$, that lies above the dotted curve corresponds to a mode with two turning points in the radial domain; hence, such a mode has a cavity that only partially fills the domain. Those frequencies that lie below the dotted line correspond to modes with only one turning point in the radial domain and that are trapped through reflection off the lower boundary.  Initially, the mode frequency increases as the azimuthal order increases until the lower turning point crosses into the domain. At larger azimuthal orders, the turning point continues to move upwards and the depth of the wave cavity shrinks commensurately. In response, the mode frequency asymptotes to a constant value that is independent of the lower boundary condition. For the prograde thermal Rossby waves, the asymptotic value can be obtained from the dispersion relation that applies for the semi-infinite domain, Equation~\eqnref{eigenfre},

\begin{equation}
    \lim_{m\to\infty} \omega_n = \frac{2\alpha \Omega}{2n+\alpha+2}
\end{equation}

\noindent For the retrograde inertial waves, the asymptotic value is zero because the angle of propagation $\chi$ approaches zero as the azimuthal order becomes large, with the result that the retrograde wave cavity shrinks to zero frequency (see Figure~\ref{fig:figure2}).

\section{Conclusion}

We have carried out a linear wave analysis for a compressible and stratified atmosphere representing a stellar convection zone rotating at a constant rate. The rotation axis is assumed to be perpendicular to the direction of stratification. By adopting an f-plane approximation, we derive and solve
dispersion relations for waves propagating through a neutrally-stable polytropic atmosphere in all three spatial directions: zonal, latitudinal and radial.

The density stratification enables radial trapping of prograde-propagating waves with frequencies
above a threshold frequency---see Equation~\eqnref{continuous_spectrum}. Low-frequency waves with frequencies below the threshold (both prograde and retrograde) cannot be trapped by an isentropic density stratification. However, the waves can reflect off of strong gradients in the buoyancy frequency (as occurs at the base of the convection zone) and thereby become radially trapped. If we consider the bottom of the convection zone to be perfectly reflective, we obtain the eigenfrequencies illustrated in Figure~\ref{fig:figure7}.  If we further place impenetrable latitudinal boundaries, as we did for our millstone shaped domain (see Figure~\ref{fig:millstone}), we obtain the eigenfrequencies shown in Figure~\ref{fig:lat_overtones}.  As expected, the eigenfrequencies generally increase as the horizontal wavenumber increases and decrease as the radial wavenumber (or radial order) increases. In particular, we point to the shape of the dispersion curves that appear in Figure~\ref{fig:lat_overtones}, as these suggest that all of the latitudinal overtones should have frequencies that initially rise as the azimuthal order increases and eventually asymptote to a common value.

While observations have yet to directly detect thermal Rossby waves, numerical simulations have long evinced such waves.  For many years now thermal Rossby waves---in their unstable, nonlinear form---have appeared as ``banana cells" or ``Busse columns."  More recently, stable long-wavelength thermal Rossby waves have been identified as well \citep{Bekki2022b}. However, only the radial and latitudinal fundamental have been reported.  The first latitudinal overtone was actually the first thermal Rossby wave to be discussed in the literature. In a linear stability analysis, \citet{Roberts1968} calculated the $\lambda=1$ thermal Rossby wave and demonstrated that this sort of wave represents the convective modes in a rotating system at convective onset. However, this antisymmetric mode turns out to be less unstable than the sectoral mode with $\lambda = 0$ and, hence, the nonlinear convective cells that appear in numerical simulations of thermal convection in a spherical shell usually possess rough symmetry across the equator.  Our calculation here may suggest the form of previously undetected tesseral modes that, due to being stable and hence low amplitude, have been skulking around in numerical simulations for many years. If observationally detected these modes can serve as seismic probes for specific entropy density.

\begin{acknowledgements}
RJ would like to thank MSRC (University of Sheffield) for partial support. BWH is supported by NASA through grants
80NSSC18K1125, 80NSSC19K0267, 80NSSC20K0193 and acknowledges collaboration with the COFFIES DSC.
\end{acknowledgements}

\appendix

\section{Nomenclature}

There is often confusion concerning the names that are applied to the different types of gravito-inertial waves, particularly thermal Rossby waves.  Beyond the fact that thermal Rossby waves have been called by a myriad of names (e.g., low-frequency prograde waves, columnar convective modes, overstable convective modes), part of the confusion arises because all of these waves are in some sense related to each other and can transition from one type of wave to another as various parameters vanish or become large.  Thermal Rossby waves are distinct from classical Rossby waves only through geometry. The restoring force is essentially the same, arising from the conservation of potential vorticity (or equivalently angular momentum).  Classical Rossby waves concern vortical motions that are largely horizontal, either because the fluid layer is thin (such as the Earth's atmosphere) or the stratification is extremely stable, thus discouraging vertical motions.  The conservation principle therefore operates on 2D spherical surfaces and this leads to retrograde propagation. Thermal Rossby waves usually reside in thick atmospheres where fluid elements are free to move vertically without inhibition. In fact, in an unstable stratification like that found in a convection zone, such motions are reinforced. Without vertical constraint, the vortex columns align instead with the rotation axis and prograde-propagating waves are produced by conservation of potential vorticity in this rotationally aligned geometry.

Even if we restrict our attention to only zonally-propagating waves in the axially constrained geometry, there are two distinct wave modes \citep[see][]{Hindman2022, Hindman2023}. In a stable stratification, with only weak rotation influence, the two gravito-inertial wave solutions consist of the retrograde and prograde branches of the internal gravity waves.  However, in an atmosphere of neutral stability, the two solutions correspond to pure inertial waves. The prograde branch now transitions to thermal Rossby waves and the retrograde branch has moved to zero frequency, becoming a stationary geostrophic mode.  Finally, in a weakly unstable stratification, both branches become prograde. The branch with the faster zonal phase speed is easily identified as thermal Rossby waves while the slow branch has been inconsistently named.  \citet{Busse1986} called this branch the thermal mode while \citet{Hindman2023} called them the slow thermal Rossby wave branch.

Here we also consider propagation latitudinally and this complicates the naming scheme further.  As we stated previously, for zonal propagation in an isentropic atmosphere, the two solution branches are prograde thermal Rossby waves and zero-frequency geostrophic modes. When the waves are allowed to propagate obliquely to the equator, the prograde branch remains a prograde inertial wave that is firmly a thermal Rossby wave.  The zero-frequency branch becomes retrograde and we choose to call it a retrograde inertial wave for lack of better choice.

Oblique propagation through a nonadiabatic stratification leads to an even further loss of clarity. The local dispersion relation reveals why,

\begin{eqnarray}
k_{z}^{2} = \left [\frac{2\Omega k_{x}}{\omega {\cal H}} + \frac{4
k_{y}^{2}\Omega^{2}}{\omega^{2}} +
k_{h}^{2}\frac{N^{2}}{\omega^{2}} \right]-  \left(k_{h}^{2} +
\frac{\omega^{2}_{ac}}{c_{s}^{2}} \right) \; .
\end{eqnarray}

\noindent There are three types of restoring forces that lead to propagation.  Stratification coupled with zonal propagation leads to the first term in the square brackets.  This term is positive only for waves with a prograde phase speed.  This term provides a compressional $\beta$-effect and leads to thermal Rossby waves.  The second term is always positive and leads to inertial waves that propagate latitudinally.  The third term arises from buoyancy and leads to internal gravity waves. Generally, however, more than one of these terms will be in operation and the wave is a three-way hybrid of internal gravity waves and the two types of inertial waves. An obvious naming scheme becomes apparent only when one, or possibly two, of the terms dominate.

\bibliography{Bibliography}

\begin{thebibliography}{}
\expandafter\ifx\csname natexlab\endcsname\relax\def\natexlab#1{#1}\fi
\providecommand{\url}[1]{\href{#1}{#1}}
\providecommand{\dodoi}[1]{doi:~\href{http://doi.org/#1}{\nolinkurl{#1}}}
\providecommand{\doeprint}[1]{\href{http://ascl.net/#1}{\nolinkurl{http://ascl.net/#1}}}
\providecommand{\doarXiv}[1]{\href{https://arxiv.org/abs/#1}{\nolinkurl{https://arxiv.org/abs/#1}}}

\bibitem[{{Abramowitz} \& {Stegun}(1968)}]{Abramowitz1968}
{Abramowitz}, M., \& {Stegun}, I.~A. 1968, {Handbook of mathematical functions
  with formulas, graphs and mathematical tables}

\bibitem[{{Bekki} {et~al.}(2022){Bekki}, {Cameron}, \& {Gizon}}]{Bekki2022b}
{Bekki}, Y., {Cameron}, R.~H., \& {Gizon}, L. 2022, \aap, 666, A135,
  \dodoi{10.1051/0004-6361/202244150}

\bibitem[{{Brun} {et~al.}(2011){Brun}, {Miesch}, \& {Toomre}}]{Brun2011}
{Brun}, A.~S., {Miesch}, M.~S., \& {Toomre}, J. 2011, \apj, 742, 79,
  \dodoi{10.1088/0004-637X/742/2/79}

\bibitem[{{Busse}(1970)}]{Busse1970}
{Busse}, F.~H. 1970, Journal of Fluid Mechanics, 44, 441,
  \dodoi{10.1017/S0022112070001921}

\bibitem[{{Busse}(1986)}]{Busse1986}
---. 1986, Journal of Fluid Mechanics, 173, 545,
  \dodoi{10.1017/S002211208600126X}

\bibitem[{{Busse} \& {Hood}(1982)}]{Busse1982}
{Busse}, F.~H., \& {Hood}, L.~L. 1982, Geophysical and Astrophysical Fluid
  Dynamics, 21, 59, \dodoi{10.1080/03091928208209005}

\bibitem[{{Christensen-Dalsgaard}(2002)}]{Christensen2002}
{Christensen-Dalsgaard}, J. 2002, Reviews of Modern Physics, 74, 1073,
  \dodoi{10.1103/RevModPhys.74.1073}

\bibitem[{{Dikpati} \& {McIntosh}(2020)}]{Dikpati2020}
{Dikpati}, M., \& {McIntosh}, S.~W. 2020, Space Weather, 18, e02109,
  \dodoi{10.1029/2019SW002109}

\bibitem[{{Gilman}(1987)}]{Gilman1987}
{Gilman}, P.~A. 1987, \apj, 318, 904, \dodoi{10.1086/165422}

\bibitem[{{Gizon} {et~al.}(2021){Gizon}, {Cameron}, {Bekki}, {Birch}, {Bogart},
  {Brun}, {Damiani}, {Fournier}, {Hyest}, {Jain}, {Lekshmi}, {Liang}, \&
  {Proxauf}}]{Gizon2021}
{Gizon}, L., {Cameron}, R.~H., {Bekki}, Y., {et~al.} 2021, \aap, 652, L6,
  \dodoi{10.1051/0004-6361/202141462}

\bibitem[{{Glatzmaier} \& {Gilman}(1981)}]{Glatzmaier1981}
{Glatzmaier}, G.~A., \& {Gilman}, P.~A. 1981, \apjs, 47, 103,
  \dodoi{10.1086/190753}

\bibitem[{{Guenther} \& {Gilman}(1985)}]{Guenther1985}
{Guenther}, D.~B., \& {Gilman}, P.~A. 1985, \apj, 295, 195,
  \dodoi{10.1086/163365}

\bibitem[{{Hanasoge} \& {Mandal}(2019)}]{Hanasoge2019}
{Hanasoge}, S., \& {Mandal}, K. 2019, \apjl, 871, L32,
  \dodoi{10.3847/2041-8213/aaff60}

\bibitem[{{Hanson} {et~al.}(2022){Hanson}, {Hanasoge}, \&
  {Sreenivasan}}]{Hanson2022}
{Hanson}, C.~S., {Hanasoge}, S., \& {Sreenivasan}, K.~R. 2022, Nature
  Astronomy, 6, 708, \dodoi{10.1038/s41550-022-01632-z}

\bibitem[{{Hathaway} \& {Upton}(2021)}]{Hathaway2021}
{Hathaway}, D.~H., \& {Upton}, L.~A. 2021, \apj, 908, 160,
  \dodoi{10.3847/1538-4357/abcbfa}

\bibitem[{{Hindman} {et~al.}(2020){Hindman}, {Featherstone}, \&
  {Julien}}]{Hindman2020}
{Hindman}, B.~W., {Featherstone}, N.~A., \& {Julien}, K. 2020, \apj, 898, 120,
  \dodoi{10.3847/1538-4357/ab9ec2}

\bibitem[{{Hindman} \& {Jain}(2022)}]{Hindman2022}
{Hindman}, B.~W., \& {Jain}, R. 2022, \apj, 932, 68,
  \dodoi{10.3847/1538-4357/ac6d64}

\bibitem[{{Hindman} \& {Jain}(2023)}]{Hindman2023}
---. 2023, \apj, 943, 127, \dodoi{10.3847/1538-4357/acaec4}

\bibitem[{{Jones} {et~al.}(2009){Jones}, {Kuzanyan}, \& {Mitchell}}]{Jones2009}
{Jones}, C.~A., {Kuzanyan}, K.~M., \& {Mitchell}, R.~H. 2009, Journal of Fluid
  Mechanics, 634, 291, \dodoi{10.1017/S0022112009007253}

\bibitem[{{Kasahara}(2003)}]{Kasahara2003}
{Kasahara}, A. 2003, Journal of Atmospheric Sciences, 60, 1085,
  \dodoi{10.1175/1520-0469(2003)60<1085:TROTHC>2.0.CO;2}

\bibitem[{{Li} {et~al.}(2000){Li}, {Finn}, {Lovelace}, \& {Colgate}}]{Li2000}
{Li}, H., {Finn}, J.~M., {Lovelace}, R.~V.~E., \& {Colgate}, S.~A. 2000, \apj,
  533, 1023, \dodoi{10.1086/308693}

\bibitem[{{Lin}(2012)}]{Lin2012}
{Lin}, M.-K. 2012, \apj, 754, 21, \dodoi{10.1088/0004-637X/754/1/21}

\bibitem[{{Lonner} {et~al.}(2022){Lonner}, {Aggarwal}, \&
  {Aurnou}}]{Lonner2022}
{Lonner}, T.~L., {Aggarwal}, A., \& {Aurnou}, J.~M. 2022, Journal of
  Geophysical Research (Planets), 127, e2022JE007356,
  \dodoi{10.1029/2022JE007356}

\bibitem[{{L{\"o}ptien} {et~al.}(2018){L{\"o}ptien}, {Gizon}, {Birch}, {Schou},
  {Proxauf}, {Duvall}, {Bogart}, \& {Christensen}}]{Loeptien2018}
{L{\"o}ptien}, B., {Gizon}, L., {Birch}, A.~C., {et~al.} 2018, Nature
  Astronomy, 2, 568, \dodoi{10.1038/s41550-018-0460-x}

\bibitem[{{Mason}(1975)}]{Mason1975}
{Mason}, P.~J. 1975, Philosophical Transactions of the Royal Society of London
  Series A, 278, 397, \dodoi{10.1098/rsta.1975.0032}

\bibitem[{{Miesch} {et~al.}(2000){Miesch}, {Elliott}, {Toomre}, {Clune},
  {Glatzmaier}, \& {Gilman}}]{Miesch2000}
{Miesch}, M.~S., {Elliott}, J.~R., {Toomre}, J., {et~al.} 2000, \apj, 532, 593,
  \dodoi{10.1086/308555}

\bibitem[{{Proxauf} {et~al.}(2020){Proxauf}, {Gizon}, {L{\"o}ptien}, {Schou},
  {Birch}, \& {Bogart}}]{Proxauf2020}
{Proxauf}, B., {Gizon}, L., {L{\"o}ptien}, B., {et~al.} 2020, \aap, 634, A44,
  \dodoi{10.1051/0004-6361/201937007}

\bibitem[{{Roberts}(1968)}]{Roberts1968}
{Roberts}, P.~H. 1968, Philosophical Transactions of the Royal Society of
  London Series A, 263, 93, \dodoi{10.1098/rsta.1968.0007}

\bibitem[{{Smith} {et~al.}(2014){Smith}, {Speer}, \& {Griffiths}}]{Smith2014}
{Smith}, C.~A., {Speer}, K.~G., \& {Griffiths}, R.~W. 2014, Journal of Physical
  Oceanography, 44, 2273, \dodoi{10.1175/JPO-D-13-0255.1}

\bibitem[{{Thuburn} {et~al.}(2002){Thuburn}, {Wood}, \&
  {Staniforth}}]{Thuburn2002}
{Thuburn}, J., {Wood}, N., \& {Staniforth}, A. 2002, Quarterly Journal of the
  Royal Meteorological Society, 128, 1793, \dodoi{10.1256/003590002320603412}

\end{thebibliography}

\propdiag
\Eigenfunction
\finitedomain
\millstone
\latovertone

\end{document}